\documentclass[runningheads]{llncs}
\usepackage{graphicx}
\usepackage{float}
\usepackage{multirow}
\usepackage{subcaption}
\usepackage{amsmath}
\usepackage{hyperref}
\usepackage{amsmath}
\usepackage[capitalize]{cleveref}
\crefname{section}{Sec.}{Secs.}
\Crefname{section}{Section}{Sections}
\Crefname{table}{Table}{Tables}
\crefname{table}{Tab.}{Tabs.}

\begin{document}
\title{Multi-Task Learning Approach for Unified Biometric Estimation from Fetal Ultrasound Anomaly Scans}

\author{
Mohammad Areeb Qazi\inst{1} *  \and
Mohammed Talha Alam\inst{1} \and
Ibrahim Almakky\inst{1} \and
Werner Gerhard Diehl\inst{2} \and
Leanne Bricker\inst{2} \and
Mohammad Yaqub \inst{1}
}
\authorrunning{ 
Qazi et al.
}
\institute{Mohamed bin Zayed University of Artificial Intelligence, \\
Abu Dhabi, United Arab Emirates
\and
Abu Dhabi Health Services Company (SEHA), \\
United Arab Emirates \\
*\email{mohammad.qazi@mbzuai.ac.ae}
}

\titlerunning{Multi-Task Learning for Fetal US}

\maketitle              
\begin{abstract}

Precise estimation of fetal biometry parameters from ultrasound images is vital for evaluating fetal growth, monitoring health, and identifying potential complications reliably. However, the automated computerized segmentation of the fetal head, abdomen, and femur from ultrasound images, along with the subsequent measurement of fetal biometrics, remains challenging. In this work, we propose a multi-task learning approach to classify the region into head, abdomen and femur as well as estimate the associated parameters. We were able to achieve a mean absolute error (MAE) of 1.08 mm on head circumference, 1.44 mm on abdomen circumference and 1.10 mm on femur length with a classification accuracy of 99.91\% on a dataset of fetal Ultrasound images. To achieve this, we leverage a weighted joint classification and segmentation loss function to train a U-Net architecture with an added classification head. The code can be accessed through \href{https://github.com/BioMedIA-MBZUAI/Multi-Task-Learning-Approach-for-Unified-Biometric-Estimation-from-Fetal-Ultrasound-Anomaly-Scans.git}{\texttt{Github}}.

\end{abstract}

\section{Introduction}

Ultrasound is a widely used non-invasive imaging modality in obstetrics for monitoring pregnancy progression and assessing fetal development. The term ``gestational age" (GA), measured in weeks, is commonly used during pregnancy to indicate the stage of pregnancy. Accurately estimating gestational age and fetal biometry parameters through ultrasound images plays a critical role in evaluating fetal growth progress and identifying any concerns during development \cite{perin2022global}. However, manual measurement of biometry parameters from ultrasound images is a time-consuming and operator-dependent task, leading to inconsistent and inaccurate measurements that can affect the interpretation of fetal growth and development \cite{joskowicz2019inter}.

Deep learning methods have demonstrated impressive performance in various medical imaging tasks, such as segmentation, registration, and classification \cite{anaya2021overview}. Despite these advancements, there remains a need to leverage deep learning techniques for the accurate estimation of gestational age and fetal biometry parameters from ultrasound images. By automating the process, deep learning models can provide consistent and reliable measurements, overcoming the limitations associated with manual estimation. Furthermore, the integration of deep learning algorithms holds promise in enabling early detection of fetal growth abnormalities, thereby improving prenatal care and the management of potential complications \cite{10.3389/fmed.2023.1098205}. The automation of biometry estimation in ultrasound imaging has emerged as a compelling area of research with the potential to enhance the efficiency of prenatal care. Clinicians recognize the significance of accurately estimating biometric parameters, including head circumference (HC), femur length (FL), and abdominal circumference (AC), in assessing fetal growth and development. As a result, there has been a growing interest in leveraging machine learning methods to automate biometry estimation, primarily focusing on HC estimation. 

The current body of literature on biometry parameter estimation primarily focuses on HC estimation, with limited attention given to other essential biometric measurements such as FL and AC. Consequently, there exists a research gap regarding the comprehensive assessment of multiple biometric parameters in ultrasound analysis \cite{Fiorentino_2023}. In this paper, we aim to fill this gap by developing an end-to-end network that incorporates the estimation of multiple biometric parameters. Instead of solely emphasizing HC estimation, our framework will incorporate the calculation of FL and AC measurements. This integrated approach, encompassing anatomical region identification and subsequent biometric parameter estimation, has the potential to be a valuable tool for clinicians in their assessment of fetal development.

In this work, our main contributions are:
\begin{enumerate}
    \item We introduce a comprehensive multi-task framework that combines classification and accurate estimation of biometric parameters, namely, HC, AC and FL.
    \item We conduct an investigation into determining the optimal weight allocations for the losses within our multi-task setup. This analysis sheds light on the crucial interplay between losses, enhancing our understanding of the system's overall performance.
\end{enumerate}

\section{Methods}

\begin{figure}[t]
    \centering
    \includegraphics[width=1.0\columnwidth]{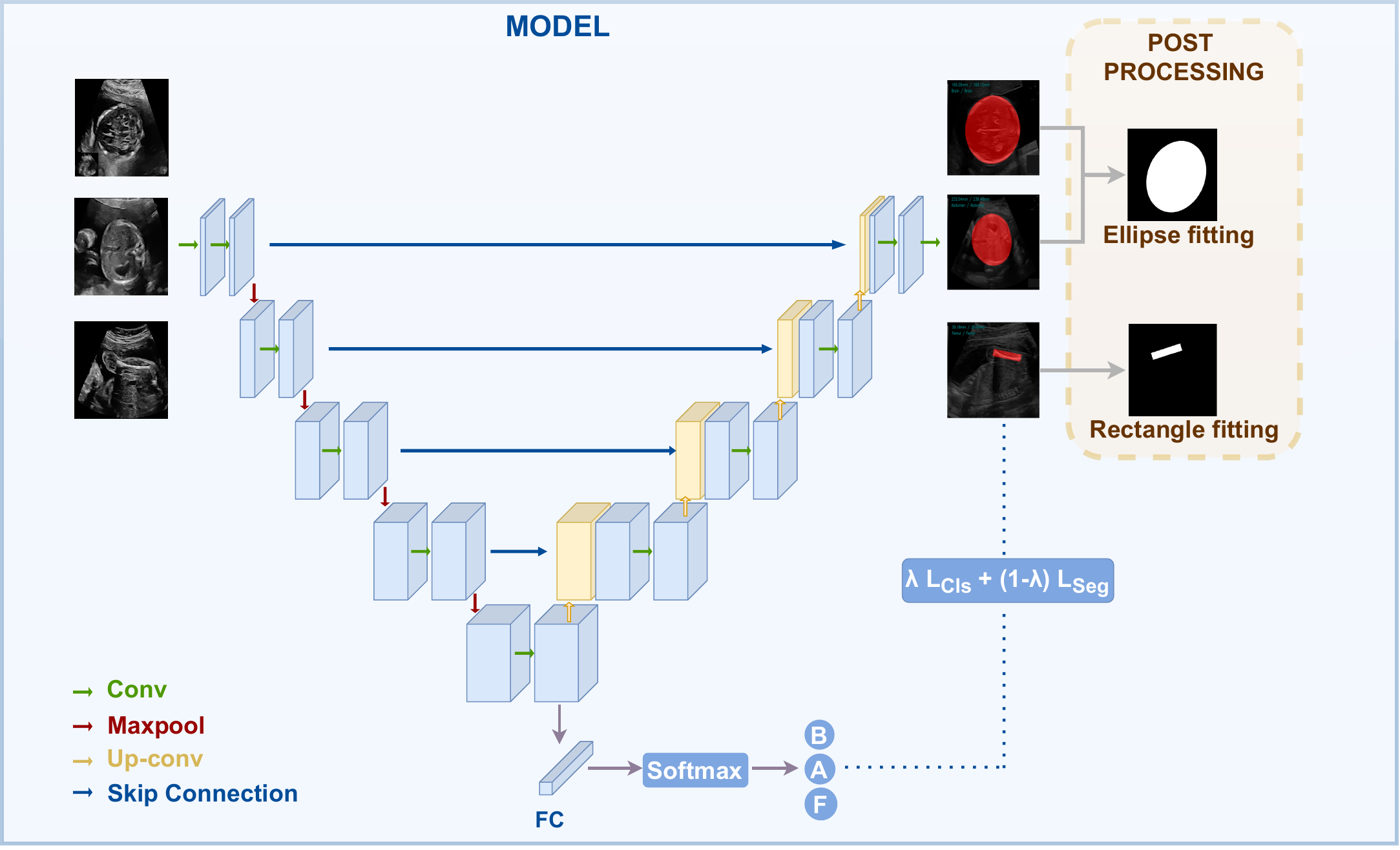}
    \caption{(a) Overview of the proposed multi-task learning network. B: Brain, A: Abdomen, F: Femur.}
    \label{framework}
\end{figure}

This research introduces a multitask learning method for 2D ultrasound images. The proposed model proposes an end-to-end Convolutional Neural Network (CNN) architecture that does classification as well as segmentation. The model predicts the class and the segmentation mask which are forwarded for post-processing, producing the bio-parameter associated with that class. The proposed framework is shown in \cref{framework}. We use U-Net \cite{10.1007/978-3-319-24574-4_28} as the backbone network for the task due to its excellent performance in 2D medical image segmentation. 

The U-Net design can be separated into three parts: an encoding path, a decoding path, and skip connections between them. U-Net incorporates stacked convolution layers in each level, as shown in \cref{framework}. In our approach, we employed five downsampling steps in the encoder to capture sufficient features. Correspondingly, the decoder used five upsampling steps to restore the features to the original size, leveraging the encoded information to predict the target's segmentation mask. Skip connections linked feature maps from the encoder to the decoder, allowing spatial information to flow and improving segmentation results. All convolution layers used 3x3 kernels, followed by batch normalization and ReLU activation. Additionally, we utilized 2x2 pooling to downsample feature maps effectively.

\subsection{Multi-task Learning Network}

Taking inspiration from the established practice in image classification, where prominent CNN models like VGG \cite{simonyan2014very} and ResNet \cite{he2016deep} incorporate high-level feature maps, our approach capitalizes on this insight. We extend the idea to our proposed multi-task learning network, enabling the extraction of shared features from U-Net for segmentation and classification. Within this network architecture, we introduce a classification branch at the base of the U-Net, as depicted in \cref{framework}. This branch utilizes feature maps from the bottleneck layer, which are flattened and directed through a fully connected (FC) layer, followed by a softmax layer. This ensemble empowers the network to predict the input image class into brain, abdomen, or femur regions, which also helps in the segmentation task. 

For the classification task, we employed the cross-entropy loss function. This loss function quantifies the dissimilarity between predicted and actual class probabilities, aiming to minimize the discrepancy during training, and it is formulated as follows:

\begin{equation}
\mathcal{L}_{\mathrm{cls}}(x, y)=-\frac{1}{N}\sum_{n=1}^{N} \log \left(\frac{\exp \left(x_{n,y_n}\right)}{\sum_{c=1}^C \exp \left(x_{n,c}\right)}\right)
\end{equation}
where $x$ represents the input, $y$ signifies the ground truth class, $C$ is the total number of classes, and $N$ is the batch size.

For the segmentation task, we employed the Dice loss function, which is formulated as follows:
\begin{equation}
\mathcal{L}_{\mathrm{seg}}(p, q) = 1 - \frac{2 \sum_i p(i) \cdot q(i)}{\sum_i p(i)^2 + \sum_i q(i)^2}
\end{equation}
where the variable $p$ signifies the ground truth binary mask, whereas $q$ stands for the predicted segmentation mask. The index $i$ pertains to the individual elements within the mask. Within our methodology, we integrate the classification loss and the segmentation loss through a linear combination, regulated by a hyperparameter denoted as $\lambda$. This combined loss, known as the multi-task loss, is precisely defined as:
\begin{equation}
    \mathcal{L}_{\mathrm{joint}} = \lambda \cdot \mathcal{L}_{\mathrm{cls}} + (1 - \lambda) \cdot \mathcal{L}_{\mathrm{seg}}
    \label{jloss}
\end{equation}

The significance of $\lambda$ lies in its capacity to assign weights to individual losses. This strategic weight allocation hinges on the concept that tasks characterized by higher training losses merit increased attention during the optimization process. By systematically varying $\lambda$ and conducting an ablation study, we gain insights into its impact and determine the optimal value that aligns with our objectives.

\subsection{Estimating the Bio-Parameters}
To calculate the bio parameter, we post-process the segmentation masks. We observed that the estimation process of HC and AC differs from the FL estimation process. This distinction arises due to the distinct shape characteristics of these anatomical regions. The head and abdomen regions can be assumed to have an elliptical shape, allowing us to apply ellipse-fitting techniques to extract the border of the head/abdomen from the segmented regions. Then the circumference can be approximated using the following equation:
\begin{align}
    circumference = \pi  \frac{a + b + 3 - (a - b)^2}{10a + 10b + \sqrt{a^2 + 14ab + b^2}}
\end{align}
where $a, b, x, y$ is the semi-minor and semi-major axis lengths and the centre coordinate of the ellipse.

In contrast to the elliptical shape of the brain and abdomen regions, we found the femur region to have a rectangular shape. To estimate the dimensions of the femur, we apply a rectangle fitting technique, allowing us to extract the length and breadth of the rectangular region. Subsequently, we calculate the perimeter of the rectangle using the obtained length and breadth measurements.

\section{Data and Experiments}

\subsection{Dataset}

Our dataset consists of images collected from a hospital over one year. The dataset comprises a total of 5,872 images from 2206 subjects, with 2,023 images depicting the brain, 2,128 images representing the abdomen, and the remaining 1,721 images focusing on the femur. To ensure comprehensive evaluation, we divided the dataset on the basis of subjects into an 80\% training set and a 20\% testing set. Additionally, we allocated 10\% of the training set as a separate validation set for fine-tuning and performance monitoring. 

The dataset included images and some associated information. Specifically, for the brain and abdomen, the dataset contained coordinates for the center, length of the the major and minor axes with an angle with the x-axis. Using these values, elliptical masks were created to represent these regions in the images. For the femur, the data had endpoints of it. These were connected to create a line with a fixed width, essentially forming a mask for the femur class. To make the data consistent, both the images and the masks were resized to a common size of 256 x 256 before using them for the model.

\subsection{Experiments}

To evaluate the effectiveness of the proposed method, we conducted ablation experiments. The results for each of these experiments are shown in \cref{Comparsion}. We started by training a separate U-Net model for each of the three classes which serves as the baseline and later helped us understand the benefit of training a single model for all three classes. Then, we added the classification head in the model and trained only the classification head by keeping the weight of the dice loss 0. Similarly, we also trained the model by keeping the classification loss weight as 0. Then we trained different models by tweaking the values of $\lambda$.

All models were implemented using Pytorch \cite{paszke2019pytorch} and were trained and tested on Quadro RTX 6000 GPUs (25G) using the ADAMax optimizer with a decaying learning rate initialized at 5e-4. The batch size and the number of epochs were set to 16 and 100 for each experiment respectively.

In this context, considering the relatively manageable nature of the classification task and the balanced dataset, we opted to utilize only accuracy as the metric for evaluating the classification component of the model. For the segmentation task, we used the biometric parameter error metric i.e. Mean Average Error (MAE). We apply the post-processing steps on the mask and then calculate MAE. MAE gauges the average magnitude of discrepancies between predicted and true values. The MAE values served as objective metrics for quantifying the results of our models in capturing these measurements. The formula for MAE is:
\begin{equation}
    MAE = \frac{1}{n} \sum_{i=1}^{n} |y_{\text{true}}^{(i)} - y_{\text{pred}}^{(i)}|
\end{equation}
Here, $n$ stands for the number of samples, $y_{\text{true}}^{(i)}$ represents the true biometric parameter value for the ith sample, and $y_{\text{pred}}^{(i)}$ is the corresponding predicted value.

\begin{table}[t]
\centering
\caption{The effects of $\lambda$ on Multi-Task Learning: The table presents an analysis of varying $\lambda$ values on the Multi-Task Learning framework. The performance metrics encompass accuracy percentages and MAE for distinct organ classes. As $\lambda$ is adjusted, notable trends emerge in both accuracy and estimation errors across the brain, abdomen, and femur classes.}
\label{Comparsion}
\resizebox{\columnwidth}{!}{%
\begin{tabular}{ccccc}
\hline
$\lambda$ & Accuracy (\%) & Brain (MAE in mm) & Abdomen (MAE in mm) & Femur (MAE in mm) \\ 
\hline
1 & 100 & - & - & - \\
0.8 & 100 & 1.50 \textpm 1.6156 & 2.33 \textpm 4.9390 & 1.50 \textpm 2.1424 \\
0.6 & 100 & 1.88 \textpm 1.7502 & 2.78 \textpm 2.9089 & 1.65 \textpm 2.6943 \\
0.4 & 99.91 & 1.39 \textpm 1.2199 & 2.08 \textpm 3.9290 & 1.53 \textpm 6.3734 \\
0.2 & 100 & 1.41 \textpm 1.2420 & 1.83 \textpm 2.0092 & 1.13 \textpm 1.6356 \\
0.1 & 100 & 1.19 \textpm 1.0825 & 1.55 \textpm 1.7488 & 1.14 \textpm 1.5085 \\
0.05 & 99.91 & 1.46 \textpm 1.2163 & 2.11 \textpm 3.7542 & 1.14 \textpm 1.7677 \\
0.025 & 100 & 1.24 \textpm 1.1018 & 1.73 \textpm 2.7895 & 1.31 \textpm 1.9683 \\
0.01 & 100 & 1.22 \textpm 1.1179 & 1.61 \textpm 1.8859 & 1.19 \textpm 1.4767 \\
0.001 & 99.91 & \textbf{1.08 \textpm 0.9702} & \textbf{1.44 \textpm 1.9269} & 1.13 \textpm 1.4498 \\
0.00001 & 99.33 & 1.18 \textpm 1.0052 & 1.54 \textpm 1.7283 & \textbf{1.10 \textpm 1.1298} \\
0 & - & 11.24 \textpm 23.1575 & 14.94 \textpm 26.8828 & 65.12 \textpm 56.0734 \\
\multicolumn{2}{c}{Separate Models} & 1.71 \textpm 1.35 & 1.55 \textpm 1.88 & 1.08 \textpm 1.18 \\ \hline
\end{tabular}%
}
\end{table}

\section{Results and Discussion}

\cref{Comparsion} provides an overview of our ablation experiment outcomes. Initially, when we used separate models for each class, we obtained MAE values of 1.71 mm for the brain, 1.55 mm for the abdomen, and 1.08 mm for the femur. We then trained only the encoder for classification across all classes (see \cref{Comparsion}, $\lambda$ = 1), successfully classifying all test images accurately. Moreover, we also trained the U-Net without the classification head (see \cref{Comparsion}, $\lambda$ = 0) to examine its effect on segmentation when classification is excluded. The results were notably worse, yielding 11.24 mm MAE for the brain, 14.95 mm MAE for the abdomen, and 65.12 mm MAE for the femur. These outcomes are considered the baseline for the ablation study, guiding us in finding the optimal balance between the weights of the two losses.

\begin{figure}[t]
    \centering
    \includegraphics[width=0.70\columnwidth]{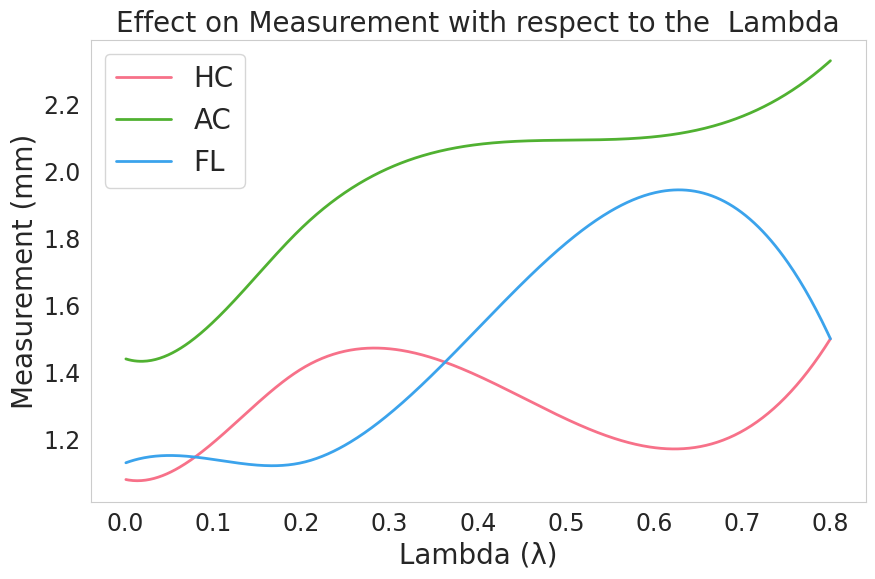}
    \caption{Effect of $\lambda$ on the Bio-Parameters: The image illustrates the impact of varying $\lambda$ values on the bio-parameters. Noticeably, the model's performance is better with lower $\lambda$ values.}
    \label{lambda_t}
\end{figure}

The value of $\lambda$ in \cref{jloss} affects the trade-off between classification and segmentation performance in multi-task learning. We experimented with various $\lambda$ values and observed interesting trends in \cref{Comparsion}. Larger $\lambda$ values gave excellent classification accuracy but had some impact on segmentation. Lowering $\lambda$ improved segmentation without significantly compromising classification. Surprisingly, very low $\lambda$ values produced good segmentation. Among all tested values, $\lambda$ = 0.001 struck the best balance. Although classification performance was consistently strong across values near 1 and smaller, the MAE saw the most improvement from 1 to 0.001. Further, lowering $\lambda$ further led to diminishing returns. \cref{lambda_t} illustrates this relationship graphically, showing how reducing $\lambda$ enhanced MAEs.

\begin{figure}[t]
    \centering
    \begin{subfigure}[b]{0.49\columnwidth}
        \centering
        \includegraphics[width=2in, height=2in]{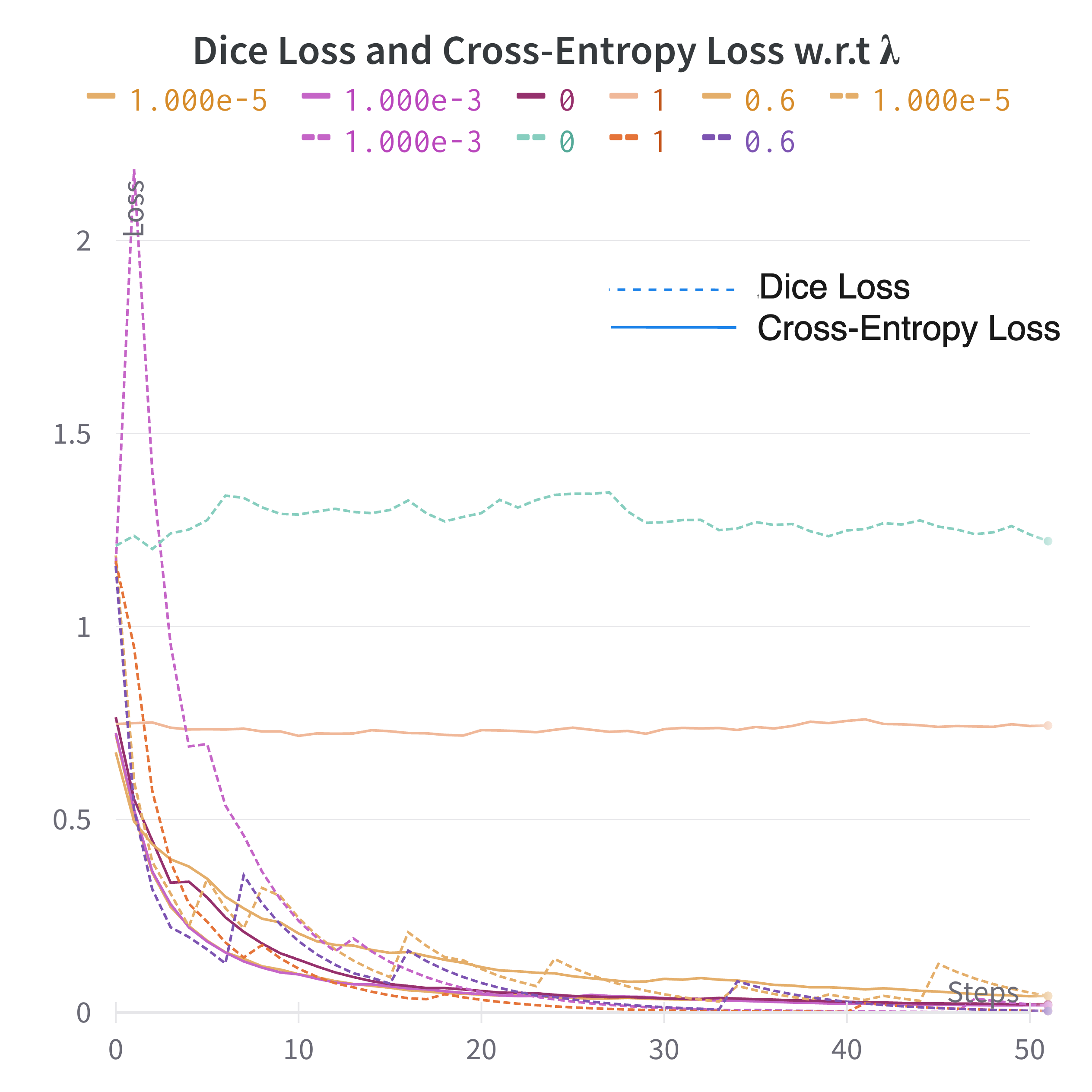}
        \caption{Dice \& Cross-Entropy Loss w.r.t $\lambda$}
        \label{cross}
    \end{subfigure}
    \hfill
    \begin{subfigure}[b]{0.46\columnwidth}
        \centering
        \includegraphics[width=2in, height=2in]{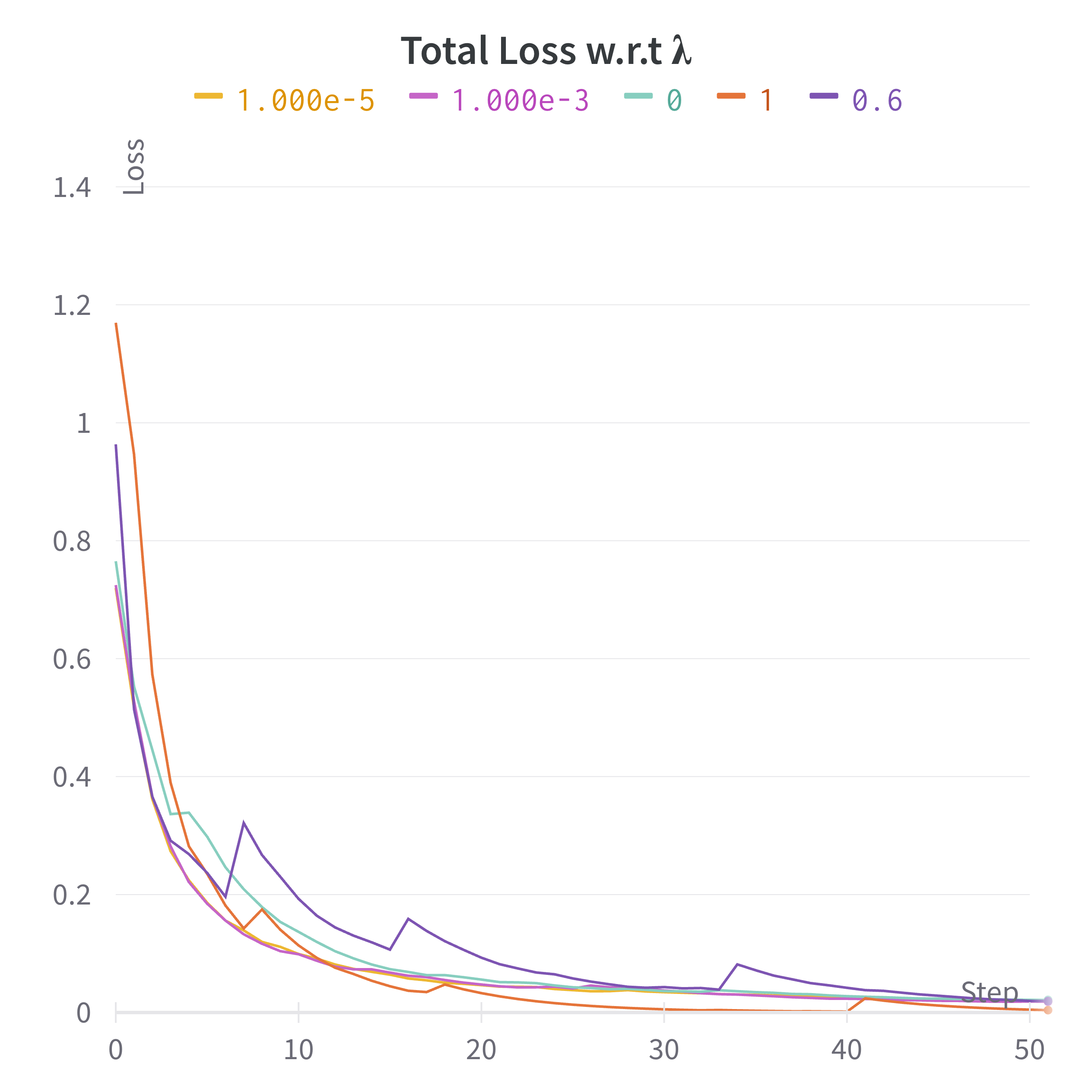}
        \caption{Total Loss w.r.t $\lambda$}
        \label{total}
    \end{subfigure}
    \caption{Training Loss Graphs with various values of $\lambda$.}
    \label{metrics}
\end{figure}

\cref{metrics} visually presents the impact of varying $\lambda$ on the convergence of different losses. In \cref{cross}, we observe that a lower $\lambda$ value led to quicker convergence of the dice loss. This can be attributed to the model's increased emphasis on the dice loss, which in turn accelerated the learning process. Conversely,higher $\lambda$ values caused greater fluctuations in the cross-entropy loss. This behavior suggests that the model found learning the classes relatively easier when the $\lambda$ was low. Consequently, the overall loss, as depicted in \cref{total}, exhibited smoother convergence when $\lambda$ was set to a smaller value. The optimal $\lambda$ value is 0.001 at which learning was notably seamless for both tasks, and the results reached their peak performance.

The pronounced improvement observed with lower values of $\lambda$ concludes that the classification task is relatively more manageable. This is attributed to the distinct structural characteristics of the classes, allowing the model to effectively grasp features and distinguish them. The multi-task learning framework plays a pivotal role in enhancing performance across individual classes, leveraging the increased volume of training data to capture a broader spectrum of variations.

The integration of a classification head yielded substantial advantages, particularly evident in improved segmentation outcomes and more precise estimations of the bio-parameter. This enhancement was prominently visible in the error reductions for specific classes: the MAE for the brain class diminished from 1.71 mm to 1.08 mm, and for the abdomen class, it decreased from 1.55 mm to 1.44 mm. Interestingly, the femur class exhibited relatively consistent performance, similar to training a standalone model. This could be attributed to the structural similarity between the brain and abdomen classes, allowing the model to leverage knowledge gained from one class to benefit the learning process for the other. 

\begin{figure}[t]
    \centering
    \begin{subfigure}{1.0\linewidth}
        \centering
        \includegraphics[width=\linewidth]{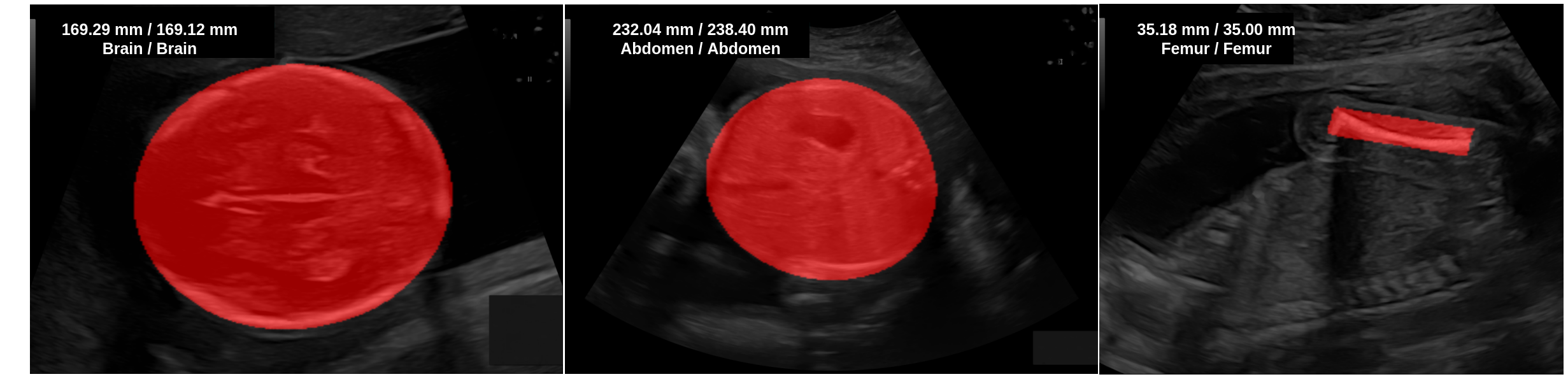}
        \caption{Success cases}
        \label{Success}
    \end{subfigure}
    \begin{subfigure}{1.0\linewidth}
        \centering
        \includegraphics[width=\linewidth]{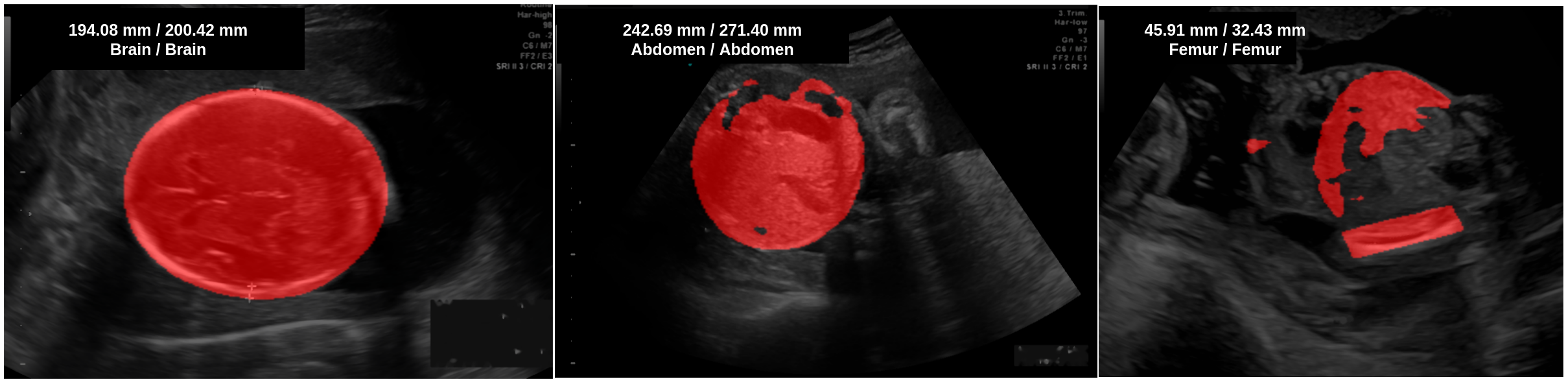}
        \caption{Failure cases}
        \label{Failure}
    \end{subfigure}
    \caption{Model Predictions: Each prediction incorporates printed estimations from the model. The prediction showcases the estimated biometric parameter (left) alongside the corresponding ground truth value (right). Further, the predicted class (left) and the actual class (right) are presented below these values.}
    \label{Out_model}
\end{figure}

Conversely, the distinct characteristics of the femur class prevented similar cross-class learning benefits. The output generated by our model when processing images from various classes is depicted in \cref{Out_model}. In cases of success, as shown in \cref{Success}, the model's predictions closely align with the ground truth, resulting in favorable Mean Average Error (MAE) values. However, challenges are evident in \cref{Failure}, which showcases instances of the model struggling with accurate segmentation, subsequently affecting the estimation. Notably, for the brain class, the segmentation mask appears sound, yet the MAE reaches around 6mm. Conversely, for the femur, complications arise due to mixed outputs with other classes, adversely impacting the predictions.

\section{Conclusion} 

In this work, an integrated approach is proposed to simultaneously classify, segment, and estimate parameters of various body organs in a fetus. The method utilizes a multi-task learning technique. By merging the cross-entropy loss with the dice loss, our model is designed to effectively estimate the parameters associated with each class. Unlike previous studies, our approach involves the calculation of multiple bio-parameters and capitalizes on insights from other classes to enhance overall performance. Additionally, we conduct experiments to evaluate the impact of varying loss weights within the multi-task learning framework. To evaluate the effectiveness of our approach, we compared it with individually trained segmentation models for each organ class. The results demonstrated noticeable improvements in terms of MAE and classification accuracy. The integrated model outperformed the individual models across evaluated organ classes. One of the key advantages of our method is its versatility, as it can be applied to multiple body organs, rather than focusing on a single organ. This offers clinicians a comprehensive tool to assess various bio-parameters simultaneously. Additionally, the computational cost of our model remains manageable, as it covers multiple organ parameters within a single framework. Future work can be done by extending this framework to more organ estimations. 

{\small
\bibliographystyle{unsrt}
\bibliography{samplepaper}
}

\end{document}


\section{Appendix}

\begin{figure}
    \centering
    \includegraphics[width=.70\columnwidth]{figures/Pre-Processing.drawio.pdf}
    \caption{Pre-Processing}
    \label{fig:PP}
\end{figure}

Our dataset consists of ultrasound scans with hard-coded labels.The initial step was to exclude all of these hard-coded labels from the input before using these scans to train our models. Given that all labels are yellow, we identified every coloured region and apply image in-painting. The measurements (FL, HC or AC) in cm or mm is displayed in a little box on the bottom-right corner of each scan. For training, these measures are not essential but for model evaluation we continue to extracted these.We utilise a commercial OCR library to find every piece of text in the scan in order to extract them. Since all labels were in the form \verb|LABEL_NAME x unit|, where \verb|unit| is either cm or mm, and \verb|LABEL_NAME| might be FL, AC, or HC. Therefore we can easily detect them by searching for \verb|LABEL_NAME| and check if the text right after it contains the text $cm$ or $mm$.

For abdomen and brain scans, we extract the ground-truth ellipse lines to create ground-truth for the segmentation task. The entire procedure is described in \cref{fig:PP}.  First, we use OCR to eliminate all the extraneous text. The dashed line remains the only area that is yellow after the text has been completely removed. Since this line is weak and not linked, we must apply picture erosion or dilation to join all the dots and make the line thicker. The ground-truth ellipse line is then obtained by fitting an ellipse to the resultant image. For the femur, we detect the two endpoints and calculate the distance between them. In order to achieve this, we must extract ground-truth important points from the given scans. In femur scans, every significant location is marked with a yellow cross pattern. So, using the pattern, we do pattern matching to retrieve the key points' positions. Then we connected the two key-point  and used the line's thickness to build a mask.

\begin{figure}
    \centering
    \includegraphics[width=1.00\columnwidth]{figures/data_preprocessing/last_updated_uNET.png}
    \caption{UNET Architecture}
    \label{fig:UNET}
\end{figure}

\begin{figure}
    \centering
    \includegraphics[width=1.00\columnwidth]{figures/data_preprocessing/last_updated_attention_uNET.png}
    \caption{Attention-UNET Architecture}
    \label{fig:A-UNET}
\end{figure}